\documentclass[twocolumn]{raa}            

\usepackage{graphicx,times}             
\usepackage{natbib}
\usepackage{amssymb,amsmath}
\bibpunct{(}{)}{;}{a}{}{,}

\usepackage[a4paper=true,dvipdfm=true,pagebackref=true]{hyperref}
\hypersetup{colorlinks = true, linkcolor = green, anchorcolor = red, citecolor = blue, filecolor = red, pagecolor = red, urlcolor = red}

\begin{document}

   \title{On the Encounter Desorption of Hydrogen Atoms on Ice Mantle
}

   \volnopage{Vol.0 (20xx) No.0, 000--000}      
   \setcounter{page}{1}          

   \author{Qiang Chang
      \inst{1}
   \and Xuli Zheng
      \inst{2}
   \and Xia Zhang
      \inst{2}
   \and Donghui Quan
      \inst{2,3}
   \and Yang Lu
      \inst{4}
   \and Qingkuan Meng
      \inst{1}
   \and Xiaohu Li
      \inst{2}
   \and Long-Fei Chen
      \inst{2}
   }

   \institute{School of Physics and Optoeletronic Engineering,
              Shandong University of Technology,
              Zibo, 255000, China; {\it changqiang@sdut.edu.cn}\\
        \and
             Xinjiang Astronomical Observatory, Chinese Academy of Sciences, 150 Science 1-Street,
             Urumqi 830011, PR China; {\it zhangx@xao.ac.cn}\\
        \and
             Department of Chemistry, Eastern Kentucky University, Richmond, KY 40475, USA\\
        \and
             School of Physics \& Astronomy, Sun Yat-Sen University, Zhuhai 519082, China\\
\vs\no
   {\small Received~~20xx month day; accepted~~20xx~~month day}}

\abstract{ At low temperatures ($\sim$ 10 K), hydrogen atoms can diffuse quickly on grain ice mantles and
frequently encounter hydrogen molecules, which cover a notable fraction of grain surface.
The desorption energy of H atoms on H$_2$ substrates is much less than that on water ice.
The H atom encounter desorption mechanism is adopted to study the enhanced desorption of H atoms on H$_2$ substrates.
Using a small reaction network, we show that the steady-state surface H abundances predicted by the rate
equation model that includes H atom encounter desorption agree reasonably well with the results from the
more rigorous microscopic Monte Carlo method. For a full gas-grain model, H atom encounter desorption can reduce
surface H abundances. Therefore, if a model adopts the encounter desorption of H atoms,
it becomes more difficult for hydrogenation products such as methanol to form,
but it is easier for C, O and N atoms to bond with each other on grain surfaces.
}

   \authorrunning{Q. Chang et al.}            
   \titlerunning{Encounter Desorption}  

   \maketitle

%
%
\section{Introduction}           
\label{sect:intro}
The grain ice mantle is observed to be mainly composed of water ice \citep{2011ApJ...740..109O},
thus, the substrate for the grain-surface chemical reactions
is usually assumed to be water ice only in the traditional astrochemical models \citep{1992ApJS...82..167H,1993MNRAS.261...83H,2006A&A...457..927G}.
However, this assumption may not be valid if the grain ice mantle is composed of other surface species
on which the desorption energies of surface species differ significantly
from these on water ice. For instance,
the desorption energy of surface species on H$_2$ substrate is more than one order of magnitude lower
than that on water ice \citep{2007ApJ...668..294C}.
If we still assume the substrates are water ice only,
hydrogen molecules may be depleted in the gas phase and are frozen on grains
at low temperatures and high densities~\citep{2015A&A...574A..24H}. However, to the best of our knowledge,
the depletion of H$_2$ has never been reported.
So, for better astrochemical modeling, there should be at least two kinds of substrates
on grain surface, water and molecular hydrogen in a chemical model.

In order to consider the much lower desorption energy of surface species on molecular hydrogen in astrochemical models,
\cite{2013MNRAS.429.3578M} made a crude approximation and assumed that the dust grains
are completely covered by H$_2$ at 10 K.  However, to the best of our knowledge, the validity of this assumption
has not been proved.
\citet{2011ApJ...735...15G} suggested another approach to solve the problem.
They used the fractional coverage of surface H$_2$ to adjust the desorption energies on water ice in their models.
The adjusted desorption energies are named the effective binding energies.
One major limitation of this approach is that an empirical modifying factor has to be used to calculate the effective binding energies.

It is straightforward to include both H$_2$ and water ice substrates in
the microscopic Monte Carlo (MC) models \citep{2005A&A...434..599C,2017SSRv..212....1C}. This numerical approach keeps
track of the trajectory of each surface species when it hops among binding sites.
If a surface species hops into a site where there is one H$_2$ molecule, the substrate for that surface species is
H$_2$ ice. Otherwise, the substrate is water ice. However, this numerical approach
is computationally  expensive when the surface temperatures are high ($\geq$ 15 K ) because the hopping events of
surface species consume most CPU time at high temperatures. Therefore, so far this approach can only be used to simulate
the chemistry of cold sources such as the dark molecular clouds.

Recently, a new mechanism called the encounter desorption was suggested so that surface H$_2$ abundances on heterogeneous grain surface
can be calculated by the rate equation (RE) approach \citep{2015A&A...574A..24H}. This mechanism considers the
desorption of hydrogen molecules when they encounter other surface H$_2$ molecules. The implementation of this mechanism
in astrochemical models is very convenient. Only one extra surface reaction gH$_2$ + gH$_2$ $\rightarrow$ H$_2$ + gH$_2$
has to be added into the chemical reaction network, where the letter g designates surface species. Moreover,
the rate coefficient of the extra surface reaction
has a simple analytical formula \citep{2015A&A...574A..24H}. The calculated surface H$_2$
abundances by the rate equation approach from models including this reaction agree reasonably well with
the results calculated by the microscopic MC method.

Hydrogen atoms can diffuse almost as fast as gH$_2$ on grain surface, so they should encounter H$_2$ frequently on grain surfaces if gH$_2$ is abundant. The desorption energy of H atoms on gH$_2$ substrates is about 45 K \citep{2007ApJ...668..294C},
so H atoms should sublime quickly when
they encounter gH$_2$ molecules just like molecular hydrogen.
Because the surface chemistry is dominated by hydrogenation reactions when the grain temperature is around 10 K,
grain surface chemistry is likely to be affected if the encounter desorption of H atoms is included in models.

In this work , we investigate how grain surface chemistry are affected by the encounter desorption of H atoms.
The paper is organized as follows. In Section 2, we derive math formula for the rate coefficient of
H atom encounter desorption on grain surface.
In Section 3, we show that for a simple reaction network that include the encounter desorption of H atoms, the surface H abundances
calculated by the rate equation approach agree reasonably well with those predicted by the microscopic MC approach.
In Section 4, we simulate a full gas-grain reaction network that includes the encounter desorption of gH$_2$
molecules and gH atoms and
present the impacts of gH atoms encounter desorption on the surface chemistry in cold cores.
In Section 5, we summarize our findings and draw the conclusions.


\section{THE ENCOUNTER DESORPTION OF H SURFACE ATOMS}
\label{sect:Encounter}

Following the gH$_2$ encounter desorption mechanism, in order to include the gH encounter desorption mechanism in a chemical model,
we can simply add an extra reaction, gH + gH$_2$ $\rightarrow$ H + gH$_2$, to the chemical reaction network.
Before deriving the rate coefficient for this reaction, we first review the encounter desorption mechanism of gH$_2$
and derive the rate coefficient for the reaction gH$_2$ + gH$_2$ $\rightarrow$ H$_2$ + gH$_2$.

\begin{figure}
	\centering
	\includegraphics[width=9cm, angle=0]{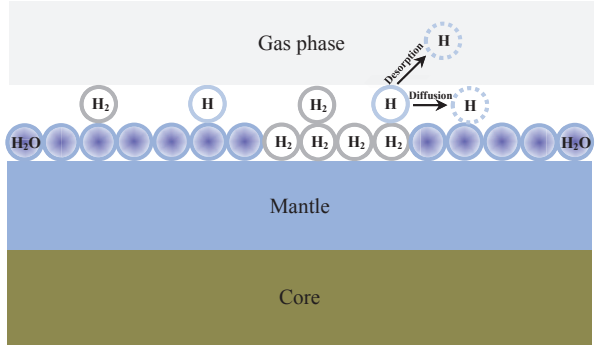}
	\caption{The diffusion of gH$_2$ and gH on substrates covered by water and gH$_2$ and the encounter desorption of gH.}
	\label{Fig1}
\end{figure}

Figure~\ref{Fig1} shows the diffusion of gH$_2$ on a heterogeneous ice mantle composed of water ice and H$_2$.
We assume that the ice mantle is mainly composed of water ice, which agrees with observations toward
cold sources \citep{2011ApJ...740..109O}.
When a hydrogen molecule hops from a water ice binding site to another site that is occupied by gH$_2$ as shown in the figure,
gH$_2$ desorption competes with the hopping events of gH$_2$.
The probability that gH$_2$ desorbs is given by the following equation \citep{2005A&A...434..599C},
 \begin{equation}
P_{H_2} = \frac{b_{H_2H_2}}{b_{H_2H_2} +k_{H_2H_2}},
\end{equation}
where $b_{H_2H_2}$ and $k_{H_2H_2}$ are the thermal desorption and hopping
rates of gH$_2$ on gH$_2$ substrates respectively.
The thermal desorption rate of gH$_2$ on gH$_2$ substrates is, $b_{H_2H_2} = \nu exp(-E_{D_{H_2H_2}}/T)$
while the hopping rate of gH$_2$ on gH$_2$ substrates is,  $k_{H_2H_2} = \nu exp(-E_{b_{H_2H_2}}/T)$,
where $E_{D_{H_2H_2}}$ and $E_{b_{H_2H_2}}$ are desorption energy and diffusion barrier of gH$_2$ on gH$_2$ substrates.
Assuming the number of binding sites occupied by gH$_2$ is much smaller than the total number of binding sites
on the grain surface, the rate coefficient for two hydrogen molecules to encounter is given by,
 \begin{equation}
r_{H_2H_2} = \frac{2k_{H_2H_2O}}{S},
\end{equation}
where $k_{H_2H_2O}$ is the hopping rate of gH$_2$ on water ice while S is the total number of binding sites on the grain surface.
Similarly, $k_{H_2H_2O} = \nu exp(-E_{b_{H_2H_2O}}/T)$  , where $E_{b_{H_2H_2O}}$ is the diffusion barrier of gH$_2$ on water ice.
Finally, we have the rate coefficient for the reaction gH$_2$ + gH$_2$ $\rightarrow$ H$_2$ + gH$_2$, which has a simple analytical form,
 \begin{equation}
r_{enc_{H_2}} = r_{H_2H_2}P_{H_2}.
\end{equation}

Figure~\ref{Fig1} also shows the diffusion of gH on grain ice mantle and the encounter of  gH and gH$_2$.
The rate coefficient for the reaction gH + gH$_2$ $\rightarrow$ H + gH$_2$, $r_{enc_{HH_2}}$ can be derived
as the follows.
First, the probability that a gH atom desorbs into the gas phase after it encounters a gH$_2$ molecule is,
 \begin{equation}
P_{HH_2} = \frac{b_{HH_2}}{b_{HH_2} +k_{HH_2}},
\end{equation}
where $b_{HH_2} = \nu exp(-E_{D_{HH_2}}/T)$ and $k_{HH_2} = \nu exp(-E_{b_{HH_2}}/T)$
are the thermal desorption and hopping rates of gH on gH$_2$ substrates respectively.
The two parameters $E_{D_{HH_2}}$ and $E_{b_{HH_2}}$ are the desorption energy and diffusion barrier of gH on gH$_2$
respectively. The gH and gH$_2$ encounter rate coefficient is,
 \begin{equation}
r_{HH_2} = \frac{k_{H_2H_2O} + k_{HH_2O}}{S},
\end{equation}
where $k_{HH_2O}$ is the hopping rate of gH on water ice, which is
dependent on the diffusion barrier of gH on water ice, $k_{HH_2O} = \nu exp(-E_{b_{HH_2O}}/T)$.
There are two possible scenarios that hydrogen atoms encounter hydrogen molecules. Either a gH atom hops into a site
already occupied by gH$_2$ or vice versa.
The encounter desorption of gH only occurs in the first case. We will explain the reason in the next section,
where the microscopic MC method will
be briefly introduced. The probability for the first case to occur is, $P = \frac{k_{HH_2O}}{k_{H_2H_2O} + k_{HH_2O}}$.
Finally we have,
 \begin{equation}
\begin{array}{lll}
r_{enc_{HH_2}} & = & r_{HH_2} P P_{HH_2}  \\
               & = & \frac{k_{HH_2O}b_{HH_2}}{S(b_{HH_2} + k_{HH_2})}
\end{array}
\end{equation}

Here we only consider thermal desorption in the above derivation. If other desorption mechanism such as photodesorption
is included in the models \citep{2007ApJ...662L..23O}, we can show that the rate coefficient for gH encounter desorption is,
 \begin{equation}
r_{enc_{HH_2}} = \frac{k_{HH_2O}(\sum_i b_{iHH_2})}{S(\sum_i b_{iHH_2} + k_{HH_2})},
\end{equation}
where $\sum_i b_{iHH_2}$ are the sum of the rates of all possible desorption mechanisms for gH on gH$_2$ substrates.

\section{NUMERICAL COMPARISON}
\label{sect:NumComp}

In order to test the validity of H atoms encounter desorption mechanism, we compare results from RE models
with the inclusion of gH and gH$_2$ encounter desorption mechanism (hereafter REED )
with these calculated  by the microscopic MC approach.
However, so far it is not possible to simulate a chemical model that include H$_2$ accretion for any meaningful length of time
using the microscopic MC method because the H$_2$ accretion and gH$_2$ hopping consume too much CPU time.
Therefore, following \citet{2015A&A...574A..24H}, we can only compare the steady-state gH abundances calculated by these two approaches.
Moreover, we also adopt a simple chemical model which only consider the accretion of H and H$_2$,
the thermal and encounter desorption of gH and gH$_2$ and the recombination of gH atoms on grain surfaces to form gH$_2$.

We adopt a typical dust grain with radius 0.1 $\mu$m. A such grain has $S =10^{6}$ binding sites.
The desorption energies of gH$_2$ on water ice and gH$_2$ are assumed to be 440 K~\citep{2015A&A...574A..24H} and 23 K~\citep{2007ApJ...668..294C} respectively
while the desorption energy of gH on water ice and gH$_2$ are assumed to be 450 K~\citep{2006A&A...457..927G} and 45 K~\citep{2007ApJ...668..294C} respectively.
The diffusion barriers are always set to be half of the desorption energy for all surface species.
Both the gas and dust temperatures are assumed to 10 K in our simulations.
We assume the density of H$_2$ to be constant in each simulation.
We perform simulations using a wide range of H nuclei densities, which vary from  2$\times$10$^4$ cm$^{-3}$
to 2$\times$10$^{14}$ cm$^{-3}$. Because the fractional abundance
of H is around 1.0$\times$10$^{-4}$ in dense clouds \citep{2007A&A...469..973C}, we assume the fractional abundance of H atoms is
fixed to be $\frac{\sqrt{2}}{4} \times 10 ^{-4}$ for all H nuclei densities for simplicty.
So, the accretion rate of H atom is,
\begin{equation}
	k_{accH}X(H)=1.0\times10^{-4} k_{accH_2}X(H_2),
\end{equation}
where $k_{accH_2}$ and $k_{accH}$ are the accretion rate coefficients of gas-phase H$_2$ and H respectively while X(H$_2$) and X(H) are the
densities of these two species respectively.
Moreover, in order for a preliminary study of the effect of the gH encounter desorption on the surface chemistry, we also use the RE approach
to calculate another chemical model which has all the above reactions other than the encounter desorption of gH atoms.
Hereafter, this model is called REED2 model.

We briefly introduce the microscopic MC approach as the follows and refer to \citet{2005A&A...434..599C} for details of this numerical method.
Grain surface binding sites form a $L\times L$ square lattice, where $L = \sqrt{S}$.
When gas-phase species accrete on grains, they are randomly located into binding sites.
At time t, a surface species gI either hop to a neighboring site or desorb after a waiting time,
\begin{equation}
 \delta t = -ln(X)/(b_I(t) + k_I(t)),
\end{equation}
where $b_I(t)$ and $k_I(t)$ are the desorption and hopping rates of species I at the time t respectively while
X is random number uniformly distributed within (0, 1).
If the site where species I resides at the time t is already occupied by gH$_2$, the value of $b_I(t)$ should be the desorption rate
of the species I on gH$_2$. Otherwise, it is the desorption rate of species I on water ice. Similarly,
the value of $k_I(t)$ depends on whether the binding site is occupied by gH$_2$ or not.
The species I hops into a neighboring site if $k_I(t) \geq Y(b_I(t) + k_I(t))$, where Y is another
random number uniformly distributed within (0, 1).
Otherwise it will desorb. A chemical reaction occurs if the species I hops into a site and encounters a reactive species.

In the microscopic MC simulations, the initial gH and gH$_2$ abundances are zero. As H and H$_2$ accrete on the grain surface,
gH abundance initially increases and then fluctuates around its mean value after the steady-state
has been reached. The average abundance of gH between time t$_0$ and t$_1$, which are both after the steady-state has been reached, is,
 \begin{equation}
	 g\bar{H}(t_0,t_1) = \frac{\sum_i i\Delta t_i}{t_1-t_0},
   \end{equation}
where $\Delta t_i$ is the time interval during which the number of gH atoms is i and $\sum_i \Delta t_i = t_1-t_0$.
If $t_1 - t_0$ is sufficiently large, $g\bar{H}$ converges to a value that is independent on $t_1$, $t_0$ nor $t_1 - t_0$.
At the low H nuclei density ($\sim 10^{-6}$ cm$^{-3}$), $t_1 - t_0$ is a time period during which about 4000 H atoms accrete
on grain surface so that $g\bar{H}$ can converge. As the H nuclei densities increase, more H atoms are required to accrete during the time
period $t_1 - t_0$ so that  $g\bar{H}$ congerges because of the larger gH population fluctuations.
We use the converged $g\bar{H}$ values to compare with the abundances of gH predicted by RE approach.

As introduced earlier, hydrogen atoms can encounter gH$_2$ in two different ways. When a gH atom hops into a site occupied by
gH$_2$, the desorption and hopping
rates of gH significantly increase because of the significantly reduced desorption energy and diffusion barrier of gH on gH$_2$.
On the other hand, when a gH$_2$ molecule hops into
a water binging site occupied by a gH atom, the gH atom cannot hop or desorb until the gH$_2$ molecule leaves the binding site
in the microscopic MC simulations \citep{2007A&A...469..973C}. After the gH$_2$ leaves the binding site, the gH atom can hop or desorb,
but its desorption energy and diffusion barrier are still these on water substrates.
Therefore, the first scenario of gH$_2$ and gH encounter must occur so that gH desorb more quickly.
If the second scenario occurs, gH$_2$ may also desorb more quickly on that site than on water ice
because the desorption energy of gH$_2$ on gH is much less than that on water ice \citep{2007ApJ...668..294C}.
A surface reaction gH + gH$_2$ $\rightarrow$ gH + H$_2$ can be included in the reaction network in the RE approach
in order to take into account of this gH$_2$ molecules desorption process when they encounter gH atoms.
However, because gH$_2$ is much more abundant than gH, this event occurs much less frequently than
the process gH$_2$ encounter other hydrogen molecules does. Therefore, in this work,
we ignore this desorption process and assume that the desorption energy of gH$_2$ on gH is the same as that on water ice.
Thus, the reaction  gH + gH$_2$ $\rightarrow$ gH + H$_2$ is not included in the reaction network.

\begin{figure}
	\centering
	\includegraphics[width=9cm, angle=0]{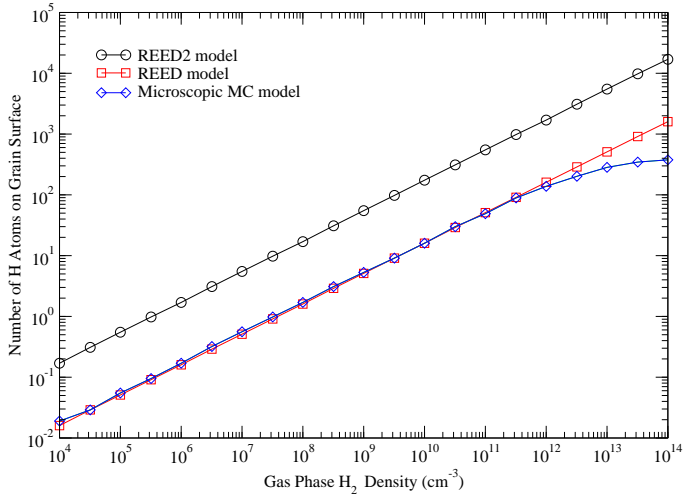}
	\caption{The steady-state surface H atom abundances with respect to H$_2$ density predicted by the REED, REED2 and microscopic
MC models using a simple chemical reaction network.
	}
	\label{Fig2}
\end{figure}

Figure~\ref{Fig2} shows the steady-state gH abundance as a function of H$_2$ density
from the microscopic MC and REED  models. It can be seen that the gH abundances predicted by the REED  models
agree very well with the results from the microscopic MC model for a wide range of H$_2$ density between
$10^4$ cm$^{-3}$ and $10^{12}$ cm$^{-3}$. At the density of $10^4$ cm$^{-3}$, gH abundance
predicted by the microscopic MC method is slightly higher than that from the REED model
because of the ``back diffusion`` problem. When the abundances of reactive surface species are very
low on grain surface, two species must visit around 3S binding sites in order to encounter
each other \citep{2006MNRAS.370.1025L,2006A&A...458..497C}.
So the reaction rates in the RE model are over estimated while the microscopic MC method already takes into
account the back diffusion. At higher H$_2$ densities, surface H$_2$ density also increases so that the
back diffusion problem becomes less significant~\citep{2006A&A...458..497C}. Thus, the REED model results agree better with
the microscopic MC model results. However when the H$_2$ density is above $10^{13}$ cm$^{-3}$, the microscopic MC model predicts that
hydrogen molecules occupy about 70 percent of the total number of grain surface binding sites \citep{2015A&A...574A..24H}.
If gH hops from one binding site to another, it is very likely that the second binding site is also occupied by gH$_2$.
This possibility is not considered by the RE approach. Therefore, the discrepancy between the
REED and microscopic model results increase as gas-phase H$_2$ abundance increases.
 At the highest H$_2$ density, the abundance of gH from the REED model is about a factor of
four larger than that from the microscopic MC model.

Figure~\ref{Fig2} also shows the steady-state gH abundance as a function of H$_2$ density for the
REED2 model. We can see that the gH encounter
desorption mechanism can decrease gH abundance by about one order of magnitude for the simple
reaction network. Grain surface reactions are dominated by hydrogenation reactions at 10 K,
so the significant decrease of gH abundance may change the abundances of icy species in a much larger gas-grain chemical model.
The effects of gH encounter desorption on the formation of surface species at 10 K in a larger chemical model
will be presented in the next section.

We fix the fractional abundance of H atoms to be  $\frac{\sqrt{2}}{4} \times 10 ^{-4}$ in the above simulations, however,
the fractional abundances of H atoms in astronomical sources may not be always around that value. In order to study
how the fractional abundances of H may affect the validity of H atoms encounter desorption mechanism, we fix the H nuclei
density to be $2\times 10^6$ cm$^{-3}$ and simulate the REED and microscopic MC models
using a wide range of H fractional abundances, which vary from $10^{-6}$ to 0.1. We found that
the gH abundances predicted by the REED model fall within 50 percent of that by the microscopic MC model over the wide
range of H fractional abundances.
Therefore, the agreement of the REED model with the microscopic MC model results is not likely
to be noticeably affected by the fractional abundance of H atoms.

\section{EFFECTS OF H ATOMS ENCOUNTER DESORPTION ON THE COLD CORE CHEMISTRY}
\label{sect:Effects}

The encounter desorption mechanism is not likely to be important when grain temperatures are high enough so that almost all
gH$_2$ molecules sublime because the encounter events would be few if gH$_2$ abundance is low. Therefore, we only
study the effects of H atoms encounter desorption on cold core chemistry.

The full gas-grain chemical reaction network was adapted from \citet{2007A&A...467.1103G} and updated based on KIDA~\citep{2011A&A...530A..61H},
which has 654 species and 6210 chemical reactions. The reactive desorption mechanism is included in this reaction network.
Its coefficient is set to be 0.01.
Initially, all species are assumed to be in the gas phase. The initial low-metal elemental abundances
are taken from \citet{2010A&A...522A..42S}.
The effect of quantum tunneling for grain-surface species is not considered in this work~\citep{1999ApJ...522..305K}.
The two-phase model is used in all simulations.
We use a standard dust grain with radius 0.1$\mu$m with surface binding site density $1.5\times10^{15}$ cm$^{-2}$.
Both the grain and gas phase temperatures are set to 10 K and the
denisty of H nuclei is $2\times 10^4$ cm$^{3}$ in our models. The
cosmic-ray ionization rate is 1.3 $\times$ 10$^{-17}$ s$^{-1}$ and the visual extinction is 10 mag.
Similar to the simple reaction network discussed above, the desorption energies of gH on water ice and gH$_2$ are 450 K and 45 K respectively
while the desorption energies of gH$_2$ on  water ice and gH$_2$ are 440 and 23 K respectively. The ratio of the diffusion barrier of
each surface species to its desorption energy is kept to be 0.5 in our models.

We simulate two chemical models.
Model A includes the reaction gH$_2$ + gH$_2$ $\rightarrow$ H$_2$ + gH$_2$ but not the reaction
gH + gH$_2$ $\rightarrow$ H + gH$_2$.
In model B, the two reactions gH$_2$ + gH$_2$ $\rightarrow$ H$_2$ + gH$_2$
and gH + gH$_2$ $\rightarrow$ H + gH$_2$ are included in its reaction network.
In total, there are 6211 and 6212 chemical reactions in models A and B respectively.
Both models are calculated by the OSU gas-grain rate equation codes~\citep{2006A&A...457..927G}.

\begin{figure}
	\centering
	\includegraphics[width=9cm, angle=0]{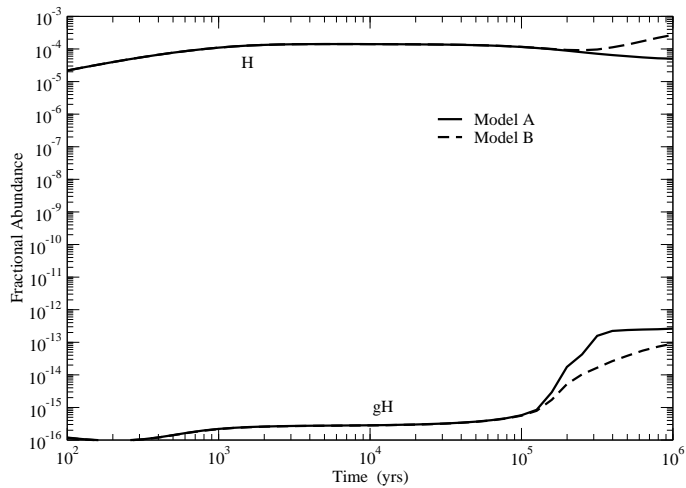}
	\caption{The gas-phase and surface H atom abundances as a function of time in models A and B.
	}
	\label{Fig3}
\end{figure}

Figure~\ref{Fig3} shows the temporal evolution of gas-phase and surface H atom abundances predicted by models A and B.
Because the H atom encounter desorption mechanism is included in model B, gH abundance from model B is typically
lower than from model A. It is interesting that at the time earlier than 10$^5$ yrs, models A and B predict
similar gH and H abundances, which can be explained as the follows. At the early time, there are abundant O, C and N in the gas phase,
so most gH atoms react with gO, gC and gN on grain surface instead of encountering gH$_2$ molecules and desorbing.
When O, C and N atoms are depleted in
the gas phase, it becomes more difficult for gH to find these atoms on grain surface to recombine, so gH can encounter
gH$_2$ molecules more frequently. As a result,
the impact of gH encounter desorption
becomes more significant. Thus, the discrepancy between the gH abundances predicted by these two models becomes larger.
At the time later than $3\times 10^5$ yrs, however, gH abundances predicted by model B become closer to that by model A,
which can be explained as the follows. In Figure~\ref{Fig3}, after $3\times 10^5$ yrs, the gas-phase H atom
abundances gradually increase with time in model B while its abundances decrease with time in model A.
At the time 10$^6$ yrs, gas-phase H abundance in model B is about a factor of 4 larger than that in model A.
Because of the higher gas-phase H atom density in model B,
more H atoms accrete on grains in model B than in model A after the time $3\times 10^5$ yrs.
So, gH abundances in model B are closer to that in model A after the time $3\times 10^5$ yrs.
The largest discrepancy between the gH abundances
predicted by these two models is about one order of magnitude, which occurs at around the time $3\times 10^5$ yrs.

\begin{figure}
	\centering
	\includegraphics[width=9cm, angle=0]{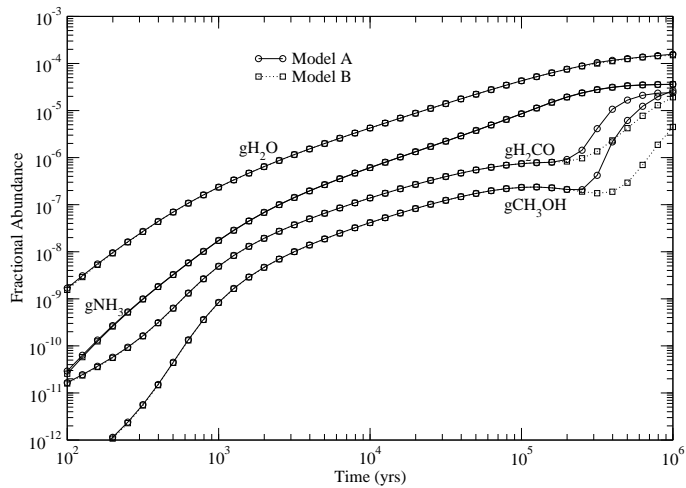}
	\caption{The fractional abundances of selected major icy species as a function of time in models A and B.
	}
	\label{Fig4}
\end{figure}

Figure~\ref{Fig4} shows selected major surface species abundances as a function of time from models A and B. These species are all
surface hydrogenated products. We can see that the production of water ice and gNH$_3$ are not much affected by the gH encounter desorption
mechanism introduced in model B. However, after the time $10^5$ yrs, model B predicts lower gCH$_3$OH and gH$_2$CO abundances than model A does
because gH atom abundances in model B are lower. Difference of the impact on surface species production
may be explained by the reaction barriers of
hydrogenation reactions. Hydrogenation reactions that form gH$_2$ and gNH$_3$ are barrierless,
thus, the rate coefficients of these hydrogenation
reactions  are large enough so that even when the gH abundances decrease, almost all gO and gN atoms can still be hydrogenated
to form water and gNH$_3$ respectively.
On the other hand, not all gCO molecules can be hydrogenated because the hydrogenation reaction gH + gCO $\rightarrow$ gHCO has a barrier.
So, when the encounter desorption is introduced in model B, less gCO can be hydrogenated, thus, less gCH$_3$OH and gH$_2$CO
molecules are formed in model B. We can also see that although it is more difficult to hydrogenate gH$_2$CO to form gCH$_3$OH in model B,
overall, the gH$_2$CO abundances are reduced by the gH encounter desorption mechanism. The abundance of another major surface species gCO in
models A and B differ by a factor less than 3. The abundances of gCO predicted by model B is always higher than that by model A
after the time $10^5$ yrs because it is easier to hydrogenate gCO in model A due to the higher gH atom abundances.

\begin{figure}
	\centering
	\includegraphics[width=9cm, angle=0]{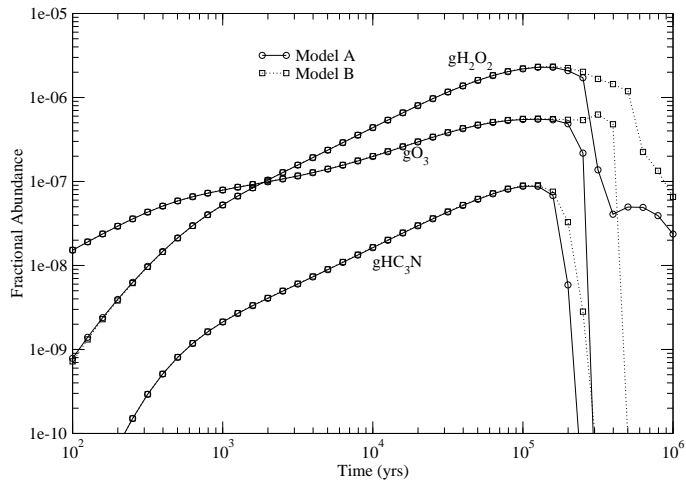}
	\caption{The temporal evolution of the fractional abundances of selected minor icy species in models A and B.
	}
	\label{Fig5}
\end{figure}

The abundances of minor surface species may be more strongly affected by the encounter desorption of gH.
Figure~\ref{Fig5} shows the temporal evolution of selected minor surface species abundances from models A and B.
After $10^5$ yrs, the abundances of these surface species from model B can be more than one order of magnitude higher
than the results from model A.
The encounter desorption of gH reduces its abundance in model B, so
it is more likely that gO, gN and gC atoms react with each other to form -O-O-, -C-C- or -C-N- chemical bonds
instead of reacting with gH atoms in this model.
Thus, more gH$_2$O$_2$, gO$_3$, and gHC$_3$N can be formed in this model than in model A.

\begin{figure}
	\centering
	\includegraphics[width=9cm, angle=0]{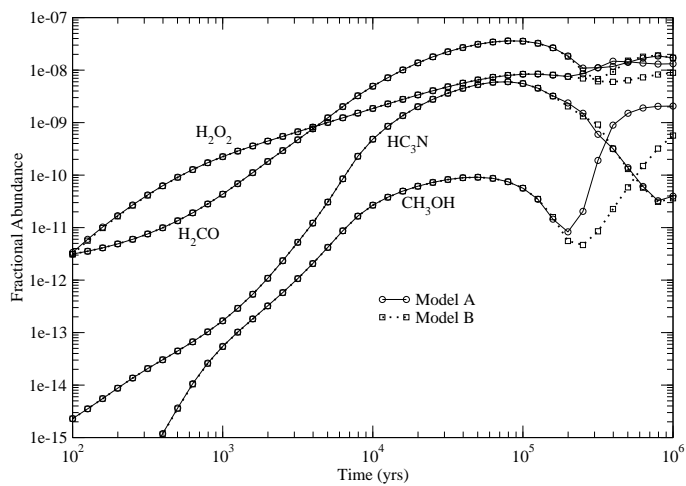}
	\caption{The fractional abundances of selected gas-phase species as a function of time in models A and B.
	}
	\label{Fig6}
\end{figure}

The encounter desorption of gH may also impact the abundances of gas-phase species. Figure~\ref{Fig6} shows the temporal evolution of selected
gaseous species abundances. If a gas-phase species is mainly produced in the gas phase, the gH encounter desorption is unlikely to impact
its abundance. For instance, HC$_3$N is mainly synthesized in gas phase, so it's abundances predicted by models A and B are almost the same.
On the other hand, methanol can hardly be produced in the gas-phase. Almost all gaseous methanol are formed via reactive desorption
on grain surface. Therefore, CH$_3$OH abundance from model B is much lower than those from model A after the time $10^5$ yrs because surface
methanol can be more efficiently synthesized in model A. Both H$_2$O$_2$ and H$_2$CO can be formed in the gas phase and via reaction desorption
on grain surfaces. So, the impact of the gH encounter desorption on their abundances is less than that on methanol abundances, but more than
that on HC$_3$N abundances.

We also performed test simulations of the above two models at a higher H nuclei density, $10^5$ cm$^{-3}$.
Similarly, model B considers the gH atom encounter desorption mechanism while model A does not.
We found that there are also less gCH$_3$OH and gH$_2$CO molecules and more surface species
such as gO$_3$ in model B  at the higher density.
On the other hand, discrepancy between species abundances in the two models becomes less significant at the higher H nuclei density.
For instance, the surface methanol abundances from the model B
are no more than a factor of 5 lower than that from model A at the higher H nuclei density while
the discrepancy can be as much as one order of magnitude at the lower H nuclei density.
We explain the decrease of the impact of the gH encounter desorption when the H nuclei density increases as the follows.
At the higher H nuclei density, more species that can react with gH accrete on grain surfaces, so it is more likely that gH atoms react with
other surface species instead of desorbing. So, the gH encounter desorption mechanism
affects grain surface chemistry less significantly at the higher H nuclei densities.

\section{CONCLUSIONS AND DISCUSSIONS}
\label{sect:Conclusion}

The significantly reduced desorption energy of hydrogen atoms when encountering hydrogen molecules on dust grains,
namely the encounter desorption mechanism of hydrogen atoms,
is investigated in this work. This process can be easily included in an astrochemical model by adding a new chemical reaction,
gH + gH$_2$ $\rightarrow$ H + gH$_2$ to the chemical reaction network. We derived the rate coefficient for this reaction. In order to
test the validity of hydrogen atom encounter desorption mechanism, we simulated a simple chemical reaction network
that includes the encounter desorption of H atoms by the RE approach and calculated the steady-state surface H atom abundances.
The calculated H atom abundances agree well with those predicted by the more rigorous microscopic MC method over a wide range of gas
phase H$_2$ and H densities. Thus, the hydrogen atom encounter desorption mechanism is a reasonable approach to take into account of
the facile desorption of hydrogen atoms on substrates composed of hydrogen molecules and water ice.

We simulated two full gas-grain chemical models under standard physical conditions that pertain to dense cores
to investigate the impact of gH encounter desorption mechanism. The gH encounter desorption mechanism is present in model B,
but not in model A.
We found that the gH atom encounter desorption can affect grain surface chemistry in two ways. Firstly, it becomes more difficult for
hydrogenation reactions with barriers to fire if the gH atom encounter desorption mechanism is included in the chemical model.
Thus, the abundances
of surface species such as methanol drops in the model because surface reactions that produce these surface species have barriers. Secondly,
gO, gN and gC atoms are more likely to react with each other instead of reacting with gH atoms in a model that includes
the gH atom encounter desorption mechanism.

Our approach can be generalized to other cases in which the facile desorption of a surface species gX on a substrate gY other than water ice
has to be considered. We can simply add one surface reaction, gX +gY  $\rightarrow$ X + gY in the chemical reaction network.
The rate coefficient for this reaction can be derived in the same way as that for gH encounter desorption.

Initially, the gH$_2$ encounter desorption mechanism was suggested to solve the problem that the coverage of gH$_2$ is too high
at low temperatures and high H nuclei densities. Our study shows that at the lower H nuclei
density, the gH encounter desorption mechanism may affect the production
of surface species significantly. On the other hand, the major advantage of this approach is that it is very simple and easy to
be implemented in a chemical model. Thus, this mechanism could be adopted widely in astrochemical modeling.

\begin{acknowledgements}	
	We thank our referee for his/her constructive comments to improve the quality of the manuscript.
The research was funded by The National Natural Science Foundation of China under grant 11673054, 11973075 and 11973099.
\end{acknowledgements}

\bibliographystyle{raa}
\bibliography{bibtex}

\begin{thebibliography}{18}
\providecommand\natexlab[1]{#1}
\providecommand\JournalTitle[1]{#1}

\bibitem[{Chang} {et~al.}(2005)]{2005A&A...434..599C}
{Chang}, Q., {Cuppen}, H.~M., \& {Herbst}, E. 2005, \aap, 434, 599

\bibitem[{Chang} {et~al.}(2006)]{2006A&A...458..497C}
{Chang}, Q., {Cuppen}, H.~M., \& {Herbst}, E. 2006, \aap, 458, 497

\bibitem[{Chang} {et~al.}(2007)]{2007A&A...469..973C}
{Chang}, Q., {Cuppen}, H.~M., \& {Herbst}, E. 2007, \aap, 469, 973

\bibitem[{Cuppen} \& {Herbst}(2007)]{2007ApJ...668..294C}
{Cuppen}, H.~M., \& {Herbst}, E. 2007, \apj, 668, 294

\bibitem[{Cuppen} {et~al.}(2017)]{2017SSRv..212....1C}
{Cuppen}, H.~M., {Walsh}, C., {Lamberts}, T., {et~al.} 2017, \ssr, 212, 1

\bibitem[{Garrod} \& {Herbst}(2006)]{2006A&A...457..927G}
{Garrod}, R.~T., \& {Herbst}, E. 2006, \aap, 457, 927

\bibitem[{Garrod} \& {Pauly}(2011)]{2011ApJ...735...15G}
{Garrod}, R.~T., \& {Pauly}, T. 2011, \apj, 735, 15

\bibitem[{Garrod} {et~al.}(2007)]{2007A&A...467.1103G}
{Garrod}, R.~T., {Wakelam}, V., \& {Herbst}, E. 2007, \aap, 467, 1103

\bibitem[{Hasegawa} \& {Herbst}(1993)]{1993MNRAS.261...83H}
{Hasegawa}, T.~I., \& {Herbst}, E. 1993, \mnras, 261, 83

\bibitem[{Hasegawa} {et~al.}(1992)]{1992ApJS...82..167H}
{Hasegawa}, T.~I., {Herbst}, E., \& {Leung}, C.~M. 1992, \apjs, 82, 167

\bibitem[{Hincelin} {et~al.}(2015)]{2015A&A...574A..24H}
{Hincelin}, U., {Chang}, Q., \& {Herbst}, E. 2015, \aap, 574, A24

\bibitem[{Hincelin} {et~al.}(2011)]{2011A&A...530A..61H}
{Hincelin}, U., {Wakelam}, V., {Hersant}, F., {et~al.} 2011, \aap, 530, A61

\bibitem[{Katz} {et~al.}(1999)]{1999ApJ...522..305K}
{Katz}, N., {Furman}, I., {Biham}, O., {Pirronello}, V., \& {Vidali}, G. 1999,
  \apj, 522, 305

\bibitem[{Lohmar} \& {Krug}(2006)]{2006MNRAS.370.1025L}
{Lohmar}, I., \& {Krug}, J. 2006, \mnras, 370, 1025

\bibitem[{Morata} \& {Hasegawa}(2013)]{2013MNRAS.429.3578M}
{Morata}, O., \& {Hasegawa}, T.~I. 2013, \mnras, 429, 3578

\bibitem[{{\"O}berg} {et~al.}(2011)]{2011ApJ...740..109O}
{{\"O}berg}, K.~I., {Boogert}, A.~C.~A., {Pontoppidan}, K.~M., {et~al.} 2011,
  \apj, 740, 109

\bibitem[{{\"O}berg} {et~al.}(2007)]{2007ApJ...662L..23O}
{{\"O}berg}, K.~I., {Fuchs}, G.~W., {Awad}, Z., {et~al.} 2007, \apjl, 662, L23

\bibitem[{Semenov} {et~al.}(2010)]{2010A&A...522A..42S}
{Semenov}, D., {Hersant}, F., {Wakelam}, V., {et~al.} 2010, \aap, 522, A42

\end{thebibliography}

\label{lastpage}

\end{document}